\documentclass[%aps,prl,
preprint,superscriptaddress,nofootinbib,preprintnumbers]{revtex4-2}
\usepackage{graphicx}% Include figure files
\usepackage{dcolumn}% Align table columns on decimal point
\usepackage{bm}% bold math
\usepackage{braket}
\usepackage{epstopdf}
\usepackage{amsmath}
\usepackage{hyperref}% add hypertext capabilities
\usepackage{xcolor}
\usepackage{cancel}
\usepackage{textcomp}
\usepackage{ulem}
\usepackage{mathtools}
\usepackage{amscd}
\usepackage{amsfonts}
\usepackage{amssymb}
\usepackage{mathrsfs}
\def\det{\operatorname{det}}

\def\={\stackrel{\bullet}{=}}

\def\({\left(}
\def\){\right)}
\def\[{\left[}
\def\]{\right]}

\def \be {\begin{equation}}
\def \ee {\end{equation}}
\def \beqa{\begin{eqnarray}}
\def \eeqa {\end{eqnarray}}
\def \beal#1 {\begin{align}#1\end{align}}
\def \bes#1 {\begin{equation}\begin{split}#1\end{split}\end{equation}}

\begin{document}

\title{
Geometric conservation in curved spacetime and entropy
}
\preprint{YITP-23-162, KEK-TH-2581, J-PARC-TH-0299
}

\author{Sinya Aoki} 
\email{saoki@yukawa.kyoto-u.ac.jp} 
\affiliation{Center for Gravitational Physics and Quantum Information, Yukawa Institute for Theoretical Physics, Kyoto University, Kitashirakawa Oiwakecho, Sakyo-ku, Kyoto 606-8502, Japan}
\author{Yoshimasa Hidaka}
\email{yoshimasa.hidaka@yukawa.kyoto-u.ac.jp}
\affiliation{Yukawa Institute for Theoretical Physics,  Kyoto University, Kyoto 606-8502, Japan
}
\affiliation{Interdisciplinary Theoretical and Mathematical Sciences Program (iTHEMS), RIKEN, Wako, Saitama 351-0198, Japan}
%\affiliation{Theory Center, Institute of Particle and Nuclear Studies, KEK, 1-1 Oho, Tsukuba 305-0801, Japan}
%\affiliation{RIKEN iTHEMS, RIKEN, Wako 351-0198, Japan}
%\affiliation{Graduate Institute for Advanced Studies, SOKENDAI, 1-1 Oho, Tsukuba, Ibaraki 305-0801, Japan}
%\affiliation{International Center for Quantum-field Measurement Systems for Studies of the Universe
%and Particles (QUP), KEK, 1-1 Oho, Tsukuba, Ibaraki 305-0801, Japan}
%\affiliation{Department of Physics, Faculty of Science, University of Tokyo, 7-3-1 Hongo Bunkyo-ku Tokyo 113-0033, Japan}

\author{Kiyoharu Kawana}
\email{kkiyoharu@kias.re.kr}
\affiliation{School of Physics, Korea Institute for Advanced Study, Seoul 02455, Korea}
\author{Kengo Shimada}
\email{kengo.shimada@yukawa.kyoto-u.ac.jp}
\affiliation{Center for Gravitational Physics and Quantum Information, Yukawa Institute for Theoretical Physics, Kyoto University, Kitashirakawa Oiwakecho, Sakyo-ku, Kyoto 606-8502, Japan}

\begin{abstract}
We provide an improved definition of new conserved quantities derived from the energy-momentum tensor in curved spacetime by introducing an additional scalar function.
We find that the conserved current and the associated conserved charge become geometric under a certain initial condition of the scalar function,
and show that such a conserved geometric current generally exists in curved spacetime.  
Furthermore, we demonstrate that the geometric conserved current agrees with the entropy current in an effective theory of a  perfect fluid,
thus the conserved charge is the total entropy of the system.  
While the geometric charge can be regarded as the entropy for a nondissipative fluid, its physical meaning should be investigated for more general cases.
\end{abstract}

\pacs{04.20.-q, 04.20.Cv, 04.70.Dy}

\maketitle

\newpage 
%%%%%%%%%%%%%%%%%%%%%%%%%%%%%%%%%%%%%
%\noindent
%{\em 1. Introduction} \hskip 0.3cm
\section{Introduction}
\label{Intro} 
In curved spacetime, while the energy-momentum tensor (EMT) is covariantly conserved, 
defining conserved energy from it remains challenging.
If the background metric is time-dependent, the energy of matter is generally not conserved.
To satisfy the total energy conservation in general relativity, one might need to abandon the covariant definition of the energy or the local definition of the energy density~\cite{Einstein:1916vd,Arnowitt:1962hi,Bondi:1962px,Komar:1962xha,Brown:1992br,Hawking:1995fd,Horowitz:1998ha,Balasubramanian:1999re,Ashtekar:1999jx}.
See Ref.~\cite{Aoki:2022gez} for a discussion on the relation between this issue and Noether's second theorem~\cite{Noether:1918zz}.

Recently, an alternative conserved current has been constructed from the EMT as $J^\mu\coloneqq \zeta T^\mu{}_\nu v^\nu$, where $T^\mu{}_\nu$ is the EMT, $\zeta$ is a scalar function, and $v^\mu$ is a given time-like unit vector field.
The associated conserved charge in some examples can be regarded as an entropy~\cite{Aoki:2020nzm}.
In addition, Noether's first theorem for the global symmetry of matter in a given curved spacetime guarantees this conservation~\cite{Aoki:2022ugd}.
In Ref.~\cite{Aoki:2020nzm}, however,  it is not clear how to choose the vector field $v^\mu$ without referring to a specific coordinate system.
 
In this paper, we refine the proposal in Ref.~\cite{Aoki:2020nzm}, and perform further analysis to elucidate the nature of this new conserved charge.
We apply the proposal in Ref.~\cite{Yokoyama:2023nld} for the perfect fluid to more general cases and choose a time-like eigenvector $u^\mu$ of the EMT for $v^\mu$. 
From the conservation condition of $J^\mu$, we determine the scalar function $\zeta$ explicitly, leading to the geometric expressions of the conserved current~\eqref{eq:J_exp} and charge~\eqref{eq:chargeQ}. 
While we implicitly use the covariant conservation of the EMT to specify properties of $u^\mu$,
$J^\mu$ is shown to be conserved without using equations of motion. Note also that our analysis can be applied not only to a background metric but also to dynamical one related to the EMT through Einstein equation.  
We show that this geometric conserved current is generic, and becomes the entropy current in the case of the perfect fluid.
Some implications of our findings are also discussed.

\section{Conserved current and conserved charge}
\label{sec:Conservation}
\subsection{%2-1. 
Decomposition of energy-momentum tensor}
We start with the EMT  in a $d$-dimensional curve spacetime given by
\begin{align}
T^\mu{}_\nu &= \varepsilon u^\mu u_\nu + P^\mu{}_\nu, 
\label{eq:EMT}
\end{align}
where $\varepsilon$ is the energy density whose eigenvector is uniquely given by a future-directed time-like unit vector 
$u^\mu$, and 
the pressure tensor $P^\mu{}_\nu$ satisfies $u_\mu P^\mu{}_\nu = P^\mu{}_\nu u^\nu=0$, so that
it is decomposed as $P_{\mu\nu} = P_{\langle \mu\nu\rangle} + h_{\mu\nu} P$ with 
$P \coloneqq  h_{\alpha\beta} P^{\alpha\beta}/(d-1)$
and $P_{\langle \mu\nu\rangle}\coloneqq   h_{\mu\alpha}  h_{\nu\beta}  ( P^{\alpha\beta} + P^{\beta\alpha})/2 - h_{\mu\nu} P$. 
We here define 
$h_{\mu\nu}\coloneqq  g_{\mu\nu}+u_\mu u_\nu$, where $g_{\mu\nu}$  is the spacetime metric with the signature $(-,+,+,\cdots,+)$,
and we do not have to specify a coordinate system for $g_{\mu\nu}$ as long as Eq.~\eqref{eq:EMT} is satisfied.
This type of EMT is a generic one at $d=4$, which describes not only massive matters but also massless ones except in special cases,
and is called the type I of the Hawking-Ellis classification~\cite{Hawking:1973uf,Martin-Moruno:2018eil}. 
In this paper, we exclusively consider this type of the EMT with the time-like $u^\mu$ at an arbitrary spacetime dimension $d$.
In addition to the EMT, which satisfies $\nabla_\mu T^\mu{}_\nu=0$, there may be $l$ conserved currents $N^\mu_i$ associated with some global symmetries as
$\nabla_\mu N_i^\mu = 0$ ($i=1,2,\cdots, l$).

\subsection{%2-2. 
Curves for \texorpdfstring{$u^\mu$}{u mu} and a family of hyper-surfaces}
We first pick up one $(d-1)$-dimensional space-like hyper-surface ${\cal H}_{d-1}$ on which the EMT is non-zero.
Therefore, $u^\mu(x)\not=0$ exists for ${}^\forall x\in {\cal H}_{d-1}$. 
Since  the hyper-surface ${\cal H}_{d-1}$ is $(d-1)$-dimensional, an arbitrary point $x_P$ in  ${\cal H}_{d-1}$ can be parametrized by a new $(d-1)$-dimensional coordinate $y^a$ ($a=1,2,\cdots, d-1$) such that ${\cal H}_{d-1} =\{  x_P^\mu (y)\, \vert\,  y\in H_{d-1} \}$,
where $H_{d-1}$ is a $(d-1)$-dimensional subspaces of ${\mathbb R}^{d-1}$. 
Note that ${\cal H}_{d-1}$ 
%$H_{d-1}$, 
may be disconnected or non-compact such that $H_{d-1}= {\mathbb R}^{d-1}$.
\begin{figure}[t]
\begin{center}
\includegraphics[width=8cm]{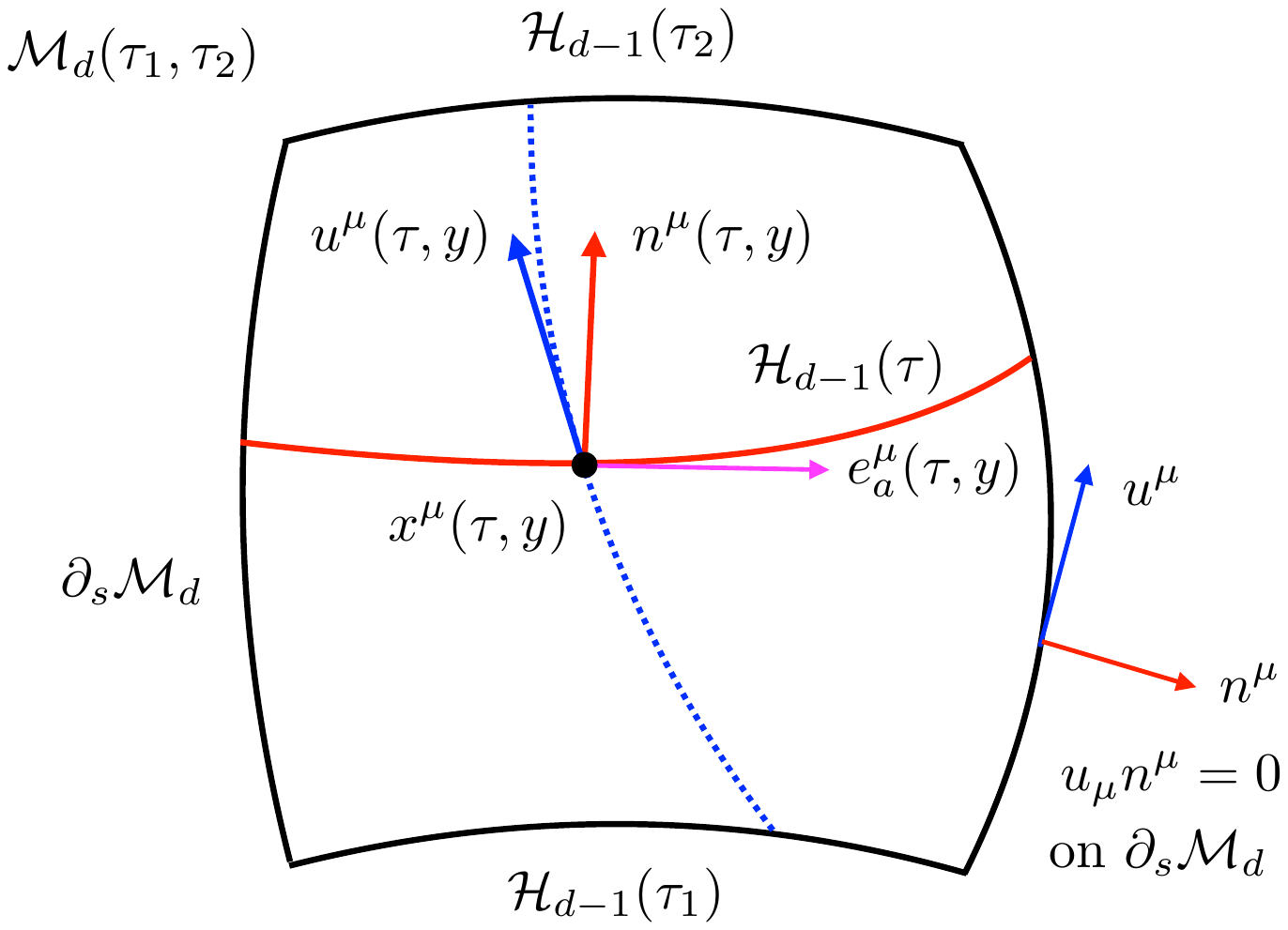}
\caption{A $d$-dimensional spacetime region ${\cal M}_d (\tau_1,\tau_2)$ and
a hyper-surface ${\cal H}_{d-1}(\tau)$ in it. 
The curves generated by the vector field $u^\mu$ are not parallel to its unit normal $n_\mu \propto -\partial_\mu \tau$ in general.
The time-like boundaries $\partial_s {\cal M}_d$ are tangent to $u^\mu$. 
For simplicity, we here plot a case with $\tilde x^0=\tau$.
}
\label{fig:hyper-surfaces}
\end{center}
\end{figure}

We introduce the timelike curve through a point $x_P(y) \in {\cal H}_{d-1}$ parametrized by $\tau$, which satisfies
\begin{equation}
\frac{d x^\mu(\tau,y)}{d \tau} = u^\mu( x^\mu(\tau,y) ),  \quad x^\mu(0,y) = x_P^\mu(y).
\label{eq:tangent}
\end{equation}
This can be integrated as
\begin{align}
x^\mu(\tau,y) &= x^\mu_P(y) + \int_0^\tau\, d\eta\, u^\mu( x(\eta,y) ) 
\label{eq:curve1}
\end{align}
for both positive and negative values of $\tau$.  
Hereafter, simplified notations such as $u^\mu(\tau,y) \coloneqq  u^\mu( x(\tau,y))$ or similar ones should be understood. 
In this paper, we assume that $\varepsilon(\tau, y)$ at $^\forall y\in H_{d-1}$ never vanish
for $\tau \in \mathbb{R}$, though it is very likely due to $\nabla_\mu T^\mu{}_\nu=0$.

We now define a family of space-like $(d-1)$-dimensional hyper-surfaces as
${\cal H}_{d-1}(\tau) \coloneqq  \left\{ x^\mu(\tau,y) \vert  {}^\exists\tau, {}^\forall y~\in H_{d-1} \right\}$,
and ${\cal H}_{d-1}(0)={\cal H}_{d-1}$ by definition.
A $d$-dimensional spacetime region  ${\cal M}_d$ associated with non-zero EMT is now given by a foliation of ${\cal H}_{d-1}(\tau)$ in terms of $\tau$.
 
So far, we do not specify a coordinate system, but it may be convenient to take a $((d-1)+1)$-decomposition $\tilde{x}^A=(\tilde x^0,\tilde x^a)$ such that $\tau=f(\tilde{x}^0)$ with $f^\prime(\tilde{x}^0) >0$, where the prime denotes the derivative with respect to $\tilde{x}^0$ and $\tilde{x}^a = y^a$ with $a = 1, \cdots, d-1$ (See Fig.~\ref{fig:hyper-surfaces}.)
Then, the metric on ${\cal M}_d$ is given by
\begin{align}
%ds^2 = 
\tilde g_{AB} d\tilde x^A d\tilde x^B &=- N^2 (d\tilde x^0)^2 + h_{ab} (d \tilde x^a + N^a d\tilde x^0) (d\tilde x^b + N^bd\tilde x^0),
\end{align}
where the induced metric on ${\cal H}_{d-1}(\tau)$ is given by $h_{ab} \coloneqq g_{\mu\nu} e^\mu_a e^\nu_b$ with $e_a^\mu \coloneqq {\partial x^\mu}/{\partial y^a}$, the shift vector by $N^a\coloneqq  h^{ab} f^\prime g_{\mu\nu} u^\mu e^\nu_b$  with $h^{ab}$ being the inverse of $h_{ab}$ , and the lapse function by $N \coloneqq \sqrt{(f^\prime)^2 + N_a N^a}$. 
In the matrix notation, we write
\begin{align}
\tilde g_{AB} =\left(
\begin{array}{cc}
  -N^2 +N_a N^a, & N_b     \\
  N_a, & h_{ab}     \\
\end{array}
\right), ~~~
\tilde g^{AB} =\frac{1}{N^2}\left(
\begin{array}{cc}
 -1 ,&   N^b     \\
   N^a, & N^2 B^{ab}    \\
\end{array}
\right), \label{eq:gtilde}
\end{align}
where 
\begin{align}
B^{ab} \coloneqq  g^{\mu\nu} \frac{\partial y^a}{\partial x^\mu}  \frac{\partial y^b}{\partial x^\nu} = h^{ab}-\frac{ N^aN^b}{N^2}. \label{eq:B}
\end{align}
In this case, 
the future-directed unit normal
to the constant $\tilde x^0$ hyper-surface ${\cal H}_{d-1}(\tau)$ is expressed as
$\tilde{n}_A = -N\delta_A^0$ while $\tilde{u}^A := u^\mu  \partial \tilde{x}^A / \partial x^\mu = \delta^A_0/f^\prime$ by definition, so that $u\cdot n = -N/f^\prime$.

For latter uses, we introduce a scalar 
$K\coloneqq g^{\mu\nu} \nabla_\mu u_\nu$, which can be expressed as
\begin{eqnarray}
    K = \partial_\mu u^\mu + u^\mu \Gamma^\nu_{~\mu \nu} = \partial_\mu u^\mu + u^\mu \partial_\mu \ln \sqrt{-g} ~,
\end{eqnarray}
where $\Gamma^\nu_{~\mu \rho}\coloneqq g^{\nu\lambda}(\partial_\mu g_{\lambda\rho}+\partial_\rho g_{\lambda\nu}-\partial_\lambda g_{\mu\rho})/2$.
In the $\tilde{x}^A$ coordinate system with the metric (\ref{eq:gtilde}), we find $\partial_A \tilde{u}^A = -  \tilde{u}^A \partial_A \ln f^\prime = - \partial_\tau \ln f^\prime$ and $\tilde{g} := \det \tilde{g}_{AB} = - N^2 h$ with $h:= \det h_{ab}$; therefore,
\begin{eqnarray}
    K= \partial_\tau \ln [ (N /f^\prime ) \sqrt{h} ] = \partial_\tau \ln [ (-n \cdot u) \sqrt{h}] ~. \label{eq:Kvu}
\end{eqnarray}

\subsection{%2-3. 
Conserved current and charge from the EMT}
As mentioned in the introduction,
the energy defined from the EMT is not conserved if the spacetime metric does not allow for the time-like Killing vector.
While one may include missing contributions from gravitational fields to make the total energy conserved,
it is difficult to do it covariantly~\cite{Aoki:2022gez}.
In this section, we instead propose an alternative conserved charge~\cite{Aoki:2020nzm}, which can be defined from the (matter) EMT alone without contribution from gravitational fields. 

We first construct a current from the EMT, a vector $v^\mu$ and a scalar function $\zeta$ as
$J^\mu (x) \coloneqq  T^\mu{}_\nu (x) \zeta(x) v^\nu(x)$,
which we require to be covariantly conserved as
$\nabla_\mu J^\mu = 0$. 
This construction of a conserved current with a scalar function $\zeta$ has been proposed previously in Ref.~\cite{Aoki:2020nzm}
\footnote{In the case of time-dependent but spherically symmetric 4-dimensional spacetime,  
the conserved current proposed in Ref.~\cite{Kodama:1979vn} agrees with $J^\mu$ in this construction if the Einstein tensor is replaced by the EMT through the Einstein equation.},
but a choice of $v^\mu$ was not explicitly given in the proposal.
In this paper, we propose $v^\mu=-u^\mu$, the time-like unit eigenvector $u^\mu$ of the EMT, which gives $J^\mu=\varepsilon(x)\zeta(x) u^\mu(x)$, and the proposal becomes coordinate independent. The use of $u^\mu$ from the perfect fluid has been proposed first in Ref.~\cite{Yokoyama:2023nld}. See also Ref.~\cite{Aoki:2022ysm}.

\subsubsection{%2-3-1. 
Determination of \texorpdfstring{$\zeta$}{zeta}}
The conservation condition applied to $J^\mu$ results in
 \begin{align}
 \nabla_\mu J^\mu =  u^\mu\partial_\mu (\zeta \varepsilon ) +\zeta \varepsilon K = \frac{\partial (\zeta\varepsilon)}{\partial \tau} +\zeta\varepsilon K = 0,
 \end{align}
  which can be solved as
 \begin{align}
 \varepsilon( \tau,y ) \zeta(\tau,y) &= \varepsilon(0,y) \zeta(0,y) \exp\left[-\int^{\tau}_{0} d\eta K( \eta,y)\right],
 \label{eq:rho-beta}
 \end{align}
where $\zeta(0,y)$ is an initial value of $\zeta$ at $\tau=0$.
Using  the expression of $K$ in Eq.~\eqref{eq:Kvu} and noticing that $(-n\cdot u)\sqrt{h} >0$ by definition, we obtain 
\begin{align}
s(\tau,y)\coloneq \zeta(\tau, y) \varepsilon (\tau, y)   
&=
\zeta(0, y)  \varepsilon (0, y)\frac{\sqrt{h} (-n\cdot u) (0, y)}{\sqrt{h} (-n\cdot u) (\tau, y)}~,
\label{eq:rho-beta2}
\end{align}
which leads to the following expression of the conserved current,
\begin{eqnarray}
J^\mu(\tau,y) &=& s(\tau,y) u^\mu(\tau,y)~. \label{eq:Jmu}
\end{eqnarray}
Note that the initial condition $ s(0,y):= \zeta(0, y)  \varepsilon (0, y)$ is controlled by  $\zeta(0, y) $.

\subsubsection{%2-3-2. 
Conserved charge}
We take a spacetime region ${\cal M}_d(\tau_1,\tau_2)$ as a foliation of ${\cal H}_{d-1}(\tau)$ at  $\tau$ between $\tau_1$ and $\tau_2$ ($\tau_1 < \tau_2$):
${\cal M}_d(\tau_1,\tau_2) \coloneqq  \{ {\cal H}_{d-1}(\tau) \  \vert \ \tau_1\le \tau \le \tau_2\}$.
Thus, $(d-1)$-dimensional boundaries of ${\cal M}_d(\tau_1,\tau_2)$ consists of two space-like hyper-surfaces, ${\cal H}_{d-1}(\tau_1)$ and ${\cal H}_{d-1}(\tau_2)$,
plus a time-like one, $\partial_s {\cal M}_d$, whose normal vector is orthogonal to $u^\mu$ everywhere on $\partial_s {\cal M}_d$, as depicted in Fig.~\ref{fig:hyper-surfaces}. 
If the parameter space $H_{d-1}$ is not simply connected, $\partial_s {\cal M}_d$ is not connected, or
$\partial_s {\cal M}_d=\emptyset$ may happen in some cases. 

Integrating  $\nabla_\mu J^\mu = 0$ over this region with Gauss's theorem, we obtain
\begin{align}
 &\int_{{\cal M}_d(\tau_1,\tau_2)}  d^dx\, \sqrt{-g} \nabla_\mu J^\mu
 = Q({\cal H}_{d-1}(\tau_2)) -Q({\cal H}_{d-1}(\tau_1)) +  \int_{\partial_s {\cal M}_d } d\Sigma_\mu J^\mu = 0,
 \label{eq:Gauss}
 \end{align}
where 
\begin{align}
 Q({\cal H}_{d-1}(\tau))&=  \int_{{\cal H}_{d-1}(\tau) }  d\Sigma_\mu\, J^\mu, 
 \label{eq:covariant}
 \end{align}
 and  $d\Sigma_\mu$ is a (hyper-)surface element on each boundary.
Since $d\Sigma_\mu J^\mu \propto d\Sigma_\mu u^\mu = 0$ on $\partial_s {\cal M}_d$, the last term in Eq.~\eqref{eq:Gauss} vanishes.
 Therefore $Q(\tau)$ is $\tau$-independent, and we thus denote $Q({\cal H}_{d-1}(\tau))=Q$.
 This charge can be understood as a kind of conserved charge derived from Noether's 1st theorem~\cite{Noether:1918zz}  in the presence of the local symmetry~\cite{Aoki:2022ugd}. 

As the charge in Eq.~\eqref{eq:covariant}
is covaraintly defined, we can express it in the $\tilde x^A$ coordinate using Eq.~\eqref{eq:Jmu} as
 \begin{align}
 Q&= \int_{{\cal H}_{d-1}(0)} d^{d-1} y \,  s(0,y) \sqrt{h} (-n\cdot u)(0,y), \label{eq:H_{d-1}(0)}
 \end{align}
 which is indeed $\tau$-independent, where we use $d \Sigma_\mu = - n_\mu \sqrt{h} d^{d-1} y$.
 One can consider infinitely many conserved charges depending on values of $\zeta$ on a specified ``initial'' time-slice.
 To be explicit, we have taken ${\cal H}_{d-1}(0)$ as this initial slice in Eq.~\eqref{eq:H_{d-1}(0)}.
 As can be seen from Fig.~\ref{fig:hyper-surfaces},
 the charge $Q$ defined on an arbitrary space-like hypersurface
instead of ${\cal H}_{d-1}(0)$ takes the same value, as long as it covers a whole region.  

\subsubsection{%2-3-3. 
A choice of an initial condition}
We now discuss a choice of $s(0,y)$, an initial condition for $s(\tau,y)$, which is scalar under the coordinate transformation of $y$. 
A simplest choice is $s(0,y)=1$, equivalent to $\zeta(0,y)=1/\varepsilon(0,y)$, which leads to a current and the corresponding charge as
\begin{eqnarray}
 J^\mu (\tau,y)&=&  \frac{\sqrt{h}(-n\cdot u)(0,y)}{\sqrt{h}(-n\cdot u)(\tau,y)}u^\mu(\tau,y),   
  \label{eq:J_exp}\\
  Q&=& \int_{{\cal H}_{d-1}(0)} d^{d-1} y \,  \sqrt{h} (-n\cdot u)(0,y).
  \label{eq:chargeQ}
\end{eqnarray}
We call the former the geometric current since it does not depend on the energy density $\varepsilon(\tau,y)$.
The charge $Q$ in \eqref{eq:chargeQ} is invariant under the coordinate transformation of $y$, thanks to a presence of $\sqrt{h}(0,y)$ in the integrand.
As we will show later, we can construct an effective theory for the perfect fluid on a curved spacetime, where
the entropy current corresponds to the geometric current $J^\mu$ defined above. 

Another choice is $s(0,y)=\varepsilon(0,y)$ ( {\it i.e.} $\zeta(0,y)=1$), whose corresponding charge becomes
\begin{eqnarray}
    Q^\prime &=&  \int_{{\cal H}_{d-1}(0)} d^{d-1} y \, \varepsilon(0,y) \sqrt{h} (-n\cdot u)(0,y).
    \label{eq:chargeQP}
\end{eqnarray}
The conserved charge $Q^\prime$, however,  is physically trivial,
since it is the total static energy (mass) at $\tau=0$ but may not be equal to the total static energy at $\tau\not=0$.
We, therefore, do not consider $Q^\prime$ in this paper.

\subsubsection{%2-3-4. 
Geometric expression of conserved currents}
As mentioned before, there may also be $l$ conserved currents $N^\mu_i$ in the system ($i=1,2,\cdots,l$), which can be written as
\begin{equation}
N^\mu_i = N_i u_i^\mu, \quad N_i:= \sqrt{-N_i\cdot N_i}, \  u_i^\mu := \frac{N_i^\mu}{N_i}.     
\end{equation}
 As in the case of $u^\mu$, we can introduce a new coordinate $(\tau_i, y_i)$ associated with $u_i^\mu$.
 The covariant conservation $\nabla_\mu N_i^\mu=0$ implies
\begin{eqnarray}
  \frac{\partial  N_i}{\partial \tau_i} +N_i K_i=0, \quad K_i := \nabla_\mu u_i^\mu, 
\end{eqnarray}
which is solved as
\begin{eqnarray}
    N_i(\tau_i,y_i) &=& N_i(0,y_i) \frac{\sqrt{h_i} (-n_i\cdot u_i) (0, y_i)}{\sqrt{h_i} (-n_i\cdot u_i) (\tau_i, y_i)},
\end{eqnarray}
and the corresponding charge, 
\begin{equation}
    Q_{i}^\prime = \int_{{\cal H}_{d-1}(\tau_i=0)} d^{d-1} y_i \, N_i(0,y_i) \sqrt{h_i} (-n_i\cdot u_i)(0,y_i).
\end{equation}
is a conserved (Noether) charge, and thus physically meaningful.

We can also define the geometric current as in Eq.~\eqref{eq:J_exp},
however, this is redundant as it can be derived as the ordinary  Noether current in this case.

 \section{%3. 
 Geometric conservation and entropy}
 In this section, we discuss the physical meaning of the geometric current \eqref{eq:J_exp}. 
 We show that the conserved current and the charge become the entropy current and the total entropy, respectively,
 in an effective theory of a  perfect fluid.
 
\subsection{%3-1. 
Geometric conservation}
To obtain the geometric conserved current \eqref{eq:J_exp}, 
 an essential property of the EMT is to provide a time-like unit vector field $u^\mu(\tau,y)$ such that
 $\varepsilon(\tau,y)\not=0$ at ${}^\forall \tau$ for $y\in H_{d-1}$ and   $\varepsilon(\tau,y)=0$  at ${}^\forall \tau$ for $y\notin H_{d-1}$,
while the detailed structure of the EMT is irrelevant. 
We here argue that the existence of such a vector field generally leads to a geometric current and charge. 

Given such a vector field $u^\mu$ satisfying the above condition,
we have the same type of decomposition as before, and accordingly, the $(d-1)$-tensor $B^{ab}$ introduced in Eq.~\eqref{eq:B}. Then, we can construct
\begin{align}
b :=\sqrt{\det B^{ab}} = \frac{f^\prime}{\sqrt{-\tilde g}}=
\frac{1}{\sqrt{h} (-n\cdot u)}~
\end{align}
%where we have used $\det B^{ab}/\det \tilde g^{AB}= \tilde g_{00} =-(f^\prime)^2$.
and
\begin{align}
J^\mu (\tau, y)  = \tilde{b} u^\mu ~,~~~ \tilde{b} := b(\tau,y)/b (0,y)  ~,
\end{align}
which is equivalent to the current given in Eq.~\eqref{eq:J_exp} and 
the covariant conservation $\nabla_\mu J^\mu = 0$ holds as before.\footnote{
One can prove the covariant conservation differently:
The inverse of $B^{ab}$ satisfies
\begin{align}
B_{ab} \partial_\mu y^a\partial_\nu y^b = g_{\mu\nu} + u_\mu u_\nu, \notag
\end{align}
since $\left(B_{ab} \partial_\mu y^a\partial_\nu y^b\right) g^{\nu\alpha} \partial_\alpha y^c = B_{ab}  B^{bc} \partial_\mu y^a =  \partial_\mu y^c$
and $\left(B_{ab} \partial_\mu y^a\partial_\nu y^b\right)  u^\nu = 0$.
Therefore,
\begin{align}
u^\alpha \nabla_\alpha b&= \frac{1}{2}  b B_{ab} u^\alpha\nabla_{\alpha} B^{ab} 
= -b g^{\mu\nu} (\nabla_\mu u^\alpha) B_{ab} \partial_\alpha y^a \partial_\nu y^b = -b g^{\mu\nu}  (\nabla_\mu u^\alpha) h_{\alpha\nu}
= - b K, \notag
\end{align}
where we have used
$u^\alpha \nabla_\alpha \partial_\mu y^a= u^\alpha \nabla_\mu \partial_\alpha y^a= -  (\nabla_\mu u^\alpha ) \partial_\alpha y^a$.
Since $u^\mu \nabla_\mu b(0,y) = 0$, we have $\nabla_\mu J^\mu  = ( u^\mu \nabla_\mu  b +  b\nabla_\mu u^\mu ) /b(0,y) =  0$. } 

To make it easier to see the conservation, we give an alternative  expression of $J^\mu$ with
the Levi-Civita tensor $\epsilon$ with $\epsilon_{01\cdots d-1}=\sqrt{-g}$:
\begin{align}
J^\mu =& - \frac{1}{(d-1)!}  \frac{1}{\sqrt{-\tilde g}} {1\over b(0,y)}
\epsilon^{\mu \alpha_1\alpha_2\cdots \alpha_{d-1}} \tilde{\epsilon}_{0a_1a_2\cdots a_{d-1}}   \partial_{\alpha_1} y^{a_1}
\partial_{\alpha_2} y^{a_2}\cdots \partial_{\alpha_{d-1}} y^{a_{d-1}}~,
\label{eq:J_alter}
\end{align}
where 
$\tilde{\epsilon}$ denotes
the Levi-Civita tensor in the $\tilde{x}^A$ coordinate.\footnote{
It can be obtained from $J^\mu =\tilde b \delta^\mu_\nu u^\nu$ with the identity
\begin{align}
\delta^\mu_\nu =& - \epsilon^{\mu \rho_1 \cdots \rho_{d-1}} \epsilon_{\nu \rho_1 \cdots \rho_{d-1}} / (d-1)!  
= - \frac{\epsilon^{\mu \rho_1 \cdots \rho_{d-1}} \partial_{\rho_1} \tilde{x}^{A_1} \cdots \partial_{\rho_{d-1}^{}} \tilde{x}^{A_{d-1}}  \tilde{\epsilon}_{A A_1 \cdots A_{d-1}}}{(d-1)! 
}  \frac{\partial \tilde{x}^{A} }{\partial x^\nu}, \notag
\end{align}
and the definition $\tilde{u}^A =  \delta^A_0$.
}
Since $\epsilon^{\mu \cdots \alpha_i \cdots \alpha_{d-1}}\nabla_\mu \partial_{\alpha_i}y^{a_i} =0$ and $J^\mu \partial_\mu y^a =0$, we get  $\nabla_\mu  J^\mu =0$.

\subsection{%3-2. 
The geometric charge for the perfect fluid}
We now consider the physical meaning of this geometric current $J^\mu$ in the case of the perfect fluid given by 
$T^\mu{}_\nu = \varepsilon u^\mu u_\nu + P (u^\mu u_\nu +\delta^\mu_\nu)$ with the pressure $P$, whose conservation leads to $\partial_\tau \varepsilon = -(\varepsilon + P) K$.
The $(d-1)$-dimensional ``comoving'' coordinate $y$ is regarded as the label of fluid element.

For simplicity,  we assume that there exists only one other conserved current $N_1^\mu$ which
is  proportional to $u^\mu$ as $N_1^\mu = N_1 u^\mu$, where $N_1$ is the corresponding charge density,
though an extension to more than one conserved current is straightforward as long as all currents are proportional to $u^\mu$.  
The corresponding conservation equation then becomes $ \nabla_\mu N_1^\mu = \partial_\tau N_1 + N_1 K = 0$.

\subsubsection{%3-2-1. 
The conservation of entropy and an effective theory} 
We define an entropy current as $s^\mu \coloneqq  s u^\mu$, where the entropy density $s$ is supposed to satisfy
thermodynamic relations, given by
$Tds = d\varepsilon - \mu_1 dN_1$ and $Ts=\varepsilon + P - \mu_1 N_1$  
with $T$ being the temperature and 
$\mu_1$ being the chemical potential for $N_1$.
Then, it is easy to see that the entropy current is conserved as
\begin{align}
\nabla_\mu s^\mu&= \frac{\partial s}{\partial \tau} + s K =  \frac{1}{T}\left( \frac{\partial \varepsilon}{\partial \tau} -\mu_1  \frac{\partial N_1}{\partial \tau}
+s T K\right) =  \frac{K}{T}\left( -\varepsilon -P +\mu_1 N_1 +s T\right) =0.
\end{align} 

Conservation of entropy in a perfect fluid can also be understood using the effective theory~\cite{Dubovsky:2005xd,Endlich:2010hf,Dubovsky:2011sj}.
The entropy current in the effective theory 
approach appears to be
only invariant under the $(d-1)$-dimensional volume-preserving diffeomorphism
because of the gauge-fixing such that
the comoving coordinate volume of a fluid element coincides with
the amount of entropy carried by that fluid element.
In this subsection, we show that, with a choice of the coordinate such that $\sqrt{h} = (-n\cdot u)^{-1}$ at $\tau = 0$ leading to $b(0,\tau) = 1$ and $\tilde b = b$, the geometric current \eqref{eq:J_exp},
\begin{eqnarray}
 J^\mu (\tau,y)&=&  \frac{1}{\sqrt{h}(-n\cdot u)(\tau,y)}u^\mu(\tau,y) = b(\tau,y) \times u^\mu(\tau,y)   ~,
  \label{eq:J_gauge_fixed}
\end{eqnarray}
corresponds to the entropy current obtained in the effective theory. Then, the total entropy is given by the corresponding charge,
\begin{eqnarray}
  Q&=& \int_{{\cal H}_{d-1}(0)} d^{d-1} y \,  .
  \label{eq:chargeQ_gauge_fixed}
\end{eqnarray}

We require that the perfect fluid should have the internal symmetries   
$y^a\rightarrow f^a(y)$ with $\det (\partial f^a /\partial y^b) = 1$ 
and $\psi\rightarrow \psi+g(y)$. 
Then, these symmetries imply that the effective action is given by a function of $b=\sqrt{\det B^{ab}}$ and $z=u^\mu \partial_\mu^{}\psi$ at the leading order of the derivative expansion. 
Using this symmetry we construct an effective theory for the perfect fluid on a curved spacetime,
whose action is given by 
\begin{equation}
S = \int d^{d}x  \sqrt{-g} F(b, z) .  
\end{equation}
The degrees of freedom are $(d-1)$-scalar fields $y^a(x)$, which describe the comoving coordinates of the fluid element and phase field $\psi(x)$.
As already seen, $J^\mu =  b u^\mu$ is covariantly conserved. 
The field  $\psi$ is a phase variable corresponding to the conserved current $N_1^\mu$, since
the action is invariant under the constant shift of $\psi$ as $\psi \to \psi + c$. The associated conserved current is obtained from the action as $N_1^\mu (x) =\partial_z F u^\mu$,
so that the charge density is expressed as $N_1 = F_z\coloneqq \partial_z F$.

\subsubsection{%3-2-2. 
The entropy as the geometric conserved charge}
We are going to show that $J^\mu$ is the entropy current for the perfect fluid, i.e., $b=s$.
The EMT is evaluated as
\begin{align}
T^{\mu\nu}(x) &\coloneqq \frac{2}{\sqrt{-g}}\frac{\partial S}{\partial g_{\mu\nu}(x)} =g^{\mu\nu} F +\frac{2F_z}{  b}\frac{\partial (J^\mu \partial_\mu \psi)}{ \partial g_{\mu\nu}}
-2(F_z z-F_{ b}   b) \frac{\partial  b}{  b \partial g_{\mu\nu}} ~,
\end{align}
where we have
\begin{align}
\frac{\partial  b }{\partial g_{\mu\nu} }&= -\frac{  b}{2} B_{ab} g^{\mu\alpha} g^{\nu\zeta} \partial_\alpha y^a\partial_\zeta y^b = -\frac{ b}{2} ( g^{\mu\nu} + u^\mu u^\nu). 
\end{align}
Since the combination $\tilde{\epsilon}_{0a_1a_2\cdots a_{d-1}}/\sqrt{-\tilde{g}}$ in Eq.~\eqref{eq:J_alter} is independent of the metric, we have $J^\mu \propto 1/\sqrt{-g}$ and
\begin{align}
\frac{\partial (J^\mu \partial_\mu \psi)}{\partial g_{\mu\nu}} =-\frac{  bz}{2} g^{\mu\nu}.   
\end{align}
Thus, the EMT becomes 
\begin{align}
T^\mu{}_\nu = (F- F_{ b}  b)\delta^\mu_\nu +(F_z z- F_{  b}   b) u^\mu u_{\nu}, %=P \delta^\mu_\nu+(\varepsilon+P)u^\mu u_\nu,
\end{align}
which leads to $ \varepsilon =F_z z-F$ and $ P=  F- F_{  b}   b$, 
in addition to $N_1=F_z$ obtained before.
These expressions give $d\varepsilon =  z d N_1 - F_{  b} d  b$, which should be consistent with $d\varepsilon=\mu_1 dN_1+T d s$.
Therefore, we find
\begin{equation}
s= b, \quad T= -F_{ b}, \quad \mu_1 =z  ~, 
\end{equation}
which correctly reproduce
thermodynamic relation as
$\varepsilon+P-\mu_1 N_1 = -F_{  b}   b = T s$. 

We therefore  conclude that the conserved geometric current %$J^\mu$
(\ref{eq:J_gauge_fixed}), and hence, its covariant form (\ref{eq:J_exp}) without gauge fixing is the entropy current as
$J^\mu = \tilde{b} u^\mu = s u^\mu$ in the case of the perfect fluid, so that the corresponding geometric conserved charge is the total entropy.
It was argued that the conserved current can be identified with the entropy current for the perfect fluid in Ref.~\cite{Aoki:2020nzm}, whose choice for the vector happens to become $u^\mu$ in this case.
In Refs.~\cite{Yokoyama:2023nld,Yokoyama:2023gxf}, the conserved current from the choice of $u^\mu$ has been explicitly constructed for the static and spherically symmetric perfect fluid and shown to become the entropy current with an appropriate choice of the temperature.

Since the entropy is conserved for the perfect fluid, or more generally, non-dissipative fluid,
it is not surprising that 
it agrees with
the total charge $Q$ in curved spacetime.\footnote{In Ref.~\cite{Sasa:2015zga}, the entropy in the flat spacetime is regarded as the Noether charge. }
To our surprise, however, the total charge $Q$ is always conserved even in cases where entropy conservation is not expected.
Currently, we do not have a good interpretation of $Q$ beyond perfect (or non-dissipative) fluid, but the formula
\begin{align}
Q& =  \int_{{\cal H}_{d-1}} \sqrt{h}(-n\cdot u) ( 0, y)\, d^{d-1}y 
\end{align}      
suggests that it measures a geometric quantity associated with the matter flux $u^\mu$, which interestingly becomes the total entropy 
for non-dissipative fluid.

In the case of general relativity, the EMT generates a curved spacetime through the Einstein equation.
Thus, the geometric charge $Q$ is the source for gravity, which we may call the gravitational charge~\cite{Aoki:2022ysm,Aoki:2023zoq}.
In the case of the non-dissipative fluid, the entropy is the source of gravity, which reminds us a statement that
the gravity is an entropic force~\cite{Jacobson:1995ab, Verlinde:2010hp}.
More we understand $Q$, more we know the nature of both entropy and gravity. 

\subsubsection{%3-2-3. 
Stefan-Boltzmann law}
In this subsection, we derive the Stefan-Boltzman law for the perfect fluid using our interpetation that $s(\tau,y)$ in \eqref{eq:Jmu} is an entropy density. For the case of the expanding Universe, see Ref.~\cite{Aoki:2023zoq}.

Assuming an equation of state (EOS)  that $P(\tau,y) =\omega(\tau,y) \varepsilon(\tau, y) $, 
together with $\partial_\tau \varepsilon= -(\varepsilon+P)K$ from the conservation of the EMT, 
we obtain
\begin{equation}
\begin{split}
\varepsilon(\tau,y) &=\varepsilon(0,y) \exp\left[-\int _0^\tau d\eta\, \{1+\omega(\tau,\eta)\} K(\eta,y)\right]~,
\end{split}
\end{equation}
where $K(\eta,y)$ is given by Eq.~(\ref{eq:Kvu}).

We now consider the time-independent EOS as $\omega(\tau,y) =\omega (y)$, which leads to
\begin{eqnarray}
\varepsilon(\tau,y) =\varepsilon(0,y)  \left( \frac{g(0,y)}{g(\tau,y)}\right)^{1+\omega(y)}~,
\end{eqnarray}
where we introduce a short-handed notation that
$g(\tau,y):= \sqrt{h}(-n\cdot u) (\tau,y)$.
The thermodynamic relation in the absence of the conserved charge such as particle numbers reads 
\begin{eqnarray}
s(\tau,y) =\frac{\varepsilon +P}{T}(\tau,y) = (1+\omega(y) )\frac{\varepsilon(\tau,y)}{T(\tau,y)} .
\end{eqnarray}
Combining these with   $s(\tau,y) =\zeta(\tau,y) \varepsilon(\tau,y)$
and the relation (\ref{eq:rho-beta2}), we obtain
\begin{eqnarray}
\varepsilon(\tau,y) =\varepsilon(0,y) \left(\frac{T(\tau,y)}{T(0,y)}\right)^{1+1/\omega(y)},  
\end{eqnarray}
which is the Stefan-Boltzmann law,
where
\begin{eqnarray}
    T(0,y) &=&\varepsilon(0,y) (1+\omega(y)).
\end{eqnarray}
Indeed $\omega(y)= 1/(d-1)=1/3$ at $d=4$, it becomes
\begin{equation}
\varepsilon(\tau,y)=\sigma_4 (y) T^4(\tau,y), \quad \sigma_4(y):= \frac{\varepsilon(0,y)}{T^4(0,y)}.
\end{equation}

This analysis also shows that $\zeta$ is proportional to the inverse temperature and 
its initial condition determines the initial temperature from  the energy density $\varepsilon$ as
\begin{equation}
    \zeta(0,y)= \frac{1}{\varepsilon(0,y)}=\frac{(1+\omega(y) )}{T(0,y)},
\end{equation}
and the time dependence of the temperature is determined in such a way that the entropy is conserved as
\begin{equation}
  T(\tau,y) = T(0,y) \left( \frac{g(0,y)}{g(\tau,y)}\right)^{\omega(y)}.   
\end{equation}

\section{%4.
Discussion}
\label{sec:Discussion}
%\noindent
%{\em 5. Discussion} \hskip 0.3cm
In this paper, we refine the proposal in Ref.~\cite{Aoki:2020nzm}, which provides a new conserved current and charge in curved spacetime, including general relativity.
With an appropriate initial condition of $\zeta$, the conserved current  become geometric as given in Eq.~\eqref{eq:J_exp}, and thus
 the conserved  charge \eqref{eq:chargeQ} just counts the total ``number'' of stream lines $u^\mu$, so that the conservation look trivial. 
 Indeed, this conserved charge becomes the number of particles for the system of massive particles interacting with each other only through gravity~\cite{Aoki:2023zoq}. 
 Surprisingly, however, the geometric conserved current agrees exactly with the entropy current in the effective theory of the perfect fluid, so that the total charge is the total entropy in the system.
 Once the vector $u^\mu$ from the EMT and $n^\mu$ and $h$ from the metric are given analytically or numerically, the entropy distribution over spacetime can be easily determined by the simple formula that $s(\tau,y)=\sqrt{h}(-n\cdot u)(0,y) /\sqrt{h}(-n\cdot u)(\tau,y)$. 

 It is important to investigate the physical meaning of the conserved current \eqref{eq:J_exp} and charge \eqref{eq:chargeQ}
 in the case of the dissipative fluid, where the entropy is generally not conserved.
The conserved charge may become trivial such as the total entropy at $\tau=0$, which however can increase at $\tau>0$.
See  $Q^\prime$ in \eqref{eq:chargeQP}, the total rest energy at $\tau=0$, for such an example.
Or the geometric conserved charge for the dissipative fluid may allow an interesting physical interpretation, which may lead to a deeper understanding of gravitational interactions.
We leave this interesting question to future studies.

\section*{Acknowledgement}
%\noindent
%{\em Acknowledgement}\hskip 0.3cm
S.A would like to thank Profs.  Matthias Blau, Yu Nakayama, Masaru Shibata, and Kenta Kiuchi for useful discussions and comments.
This work is supported in part by the Grant-in-Aid of the Japanese Ministry of Education, Sports, Culture, Sciences and Technology (MEXT) for Scientific Research (No.~JP22H00129). 
The work of K.K. is supported by KIAS Individual Grants, Grant No. 090901.

\bibliographystyle{utphys}
\bibliography{refer}

\end{document}